\documentstyle[aps,epsfig]{revtex}

\def\Journal#1#2#3#4{{#1} {\bf #2}, #3 (#4)}

\def\NPB{{\em Nucl. Phys.} B}
\def\NPA{{\em Nucl. Phys.} A}
\def\PLB{{\em Phys. Lett.} B}
\def\PRL{\em Phys. Rev. Lett.}
\def\PRD{{\em Phys. Rev.} D}
\def\PRC{{\em Phys. Rev.} C}

\begin{document}

\preprint{\vbox {\hspace*{\fill} DOE/ER/40762-172\\
                 {\fill} UMD PP\#99-078}} 
\vspace{.5in}

\title {Testing  Low Energy Theorems in Nucleon-Nucleon Scattering}

\author{Thomas D. Cohen}

\address{Department of 
Physics, University of~Maryland, College~Park, MD~20742-4111}

\author{James M. Hansen}

\address{Montgomery Blair  High School,  Silver Spring, MD 20901}

\maketitle

\vspace{.25in}

\begin{abstract}

Low energy theorems have been derived for the coefficients of the effective
range expansion in s-wave nucleon-nucleon scattering valid to leading nontrivial
order in an expansion based  $Q$ counting, a scheme in which both 
$m_\pi$ and $1/a$ (where $a$ is the scattering length) are treated as small mass scales.  Previous tests of these theorems based on coefficients extracted from
scattering data indicate a pattern of gross violations which suggested
serious problems for the perturbative treatment of pions implicit in
 $Q$ counting.  We discuss the possibility
that uncertainties associated with extracting 
the coefficients from the scattering data make such tests invalid.  
Here we  show that errors in 
 the s-wave phase shift extractions are sufficiently small to test
 direct test predictions from $Q$ counting at next to
leading order.  In particular we show that there exist low energy theorems 
for the sum of all terms in the effective range expansion beyond the first two which allow for precise tests.  These low energy theorems fail badly
which suggests that pionic aspects of $Q$ counting are not under control.

\vspace{.5in}
\end{abstract}

\section{Introduction}
There has been considerable interest in the use of effective field theory
(EFT) techniques in nuclear physics during the past several years
\cite{Weinberg1,KoMany,Parka,KSWa,CoKoM,DBK,cohena,Scal,Fria,Sa96,LMa,GPLa,Adhik,RBMa,Bvk,aleph,Parkb,Gegelia1,Gegelia2,steelea,KSW,KSW2,CGSS,cohenb,Sav,CH,CH2,MS1,MS2,steeleb}.  Much of the  goal of this work   is to use power counting 
ideas associated with chiral symmetry to nuclear physics.  This is not 
simple since apart from $m_\pi$, the inverse s-wave scattering length, $1/a$
is another light scale in the problem.  Many of the approaches
 beginning with Weinberg's\cite{Weinberg1}
formulate the expansion at the level of a  two-particle 
irreducible kernel  rather than for
observables.  While such an approach provides an organizing principle for
 calculations,  it provides no systematic 
estimate of the accuracy of particular observables in terms of power counting.
 Recently a scheme was introduced in which observables can be
expressed in terms of a consistent power counting scheme\cite{KSW,CH,CH2,MS1,MS2}.

This scheme is  based on power counting in a single scale, Q
\begin{equation}
m_\pi \sim   Q  \; \; \; 
1/a \sim  Q \; \; \; 
k \sim  Q
\label{pc}
\end{equation}
In this power counting, all other scales are assumed to be heavy and will 
collectively be symbolized by $\Lambda$.
This power counting scheme  describes low momentum physics in that $k/ \Lambda \ll 1$.  There can be rapid momentum dependence of some observables, however,
since the expansion for any observable includes all orders in $k a$ and $k/m_\pi$.   This power counting scheme has been
implemented using dimensional regularization\cite{KSW,MS1,MS2} and
 directly in configuration space using a
cutoff \cite{CH}. We note in passing the fact that $1/a$  is formally treated as being of
the same order as $m_\pi$ and $k$ is not emphasized in the original papers
of Kaplan, Savage and Wise\cite{KSW,KSW2}.   It is implicit, however, in 
the expression for the leading order ($Q^{-1}$) amplitude which is given by
$-4\pi/[M \, (1/a + i k)]$.  Note that if $k$ and $1/a$ were of different orders
one could expand out the denominators.  One test that the $Q$ counting formally 
involves treating  $1/a$ as being of the same order is found in the cutoff
treatment of ref. \cite{CH} where the rules in eq.~(\ref{pc}) were explicitly
 used to  derive an expression for the phase shifts which is formally
 equivalent to  the  expressions derived by Kaplan, Savage and Wise\cite{KSW,KSW2}.

In a previous  paper\cite{CH2} we used $Q$ counting   to derive low energy theorems for
coefficients of the effective range expansion (ERE) at leading nontrivial order in $Q$ counting.   
 The ERE is a parameterization of s-wave scattering given by
\begin{equation} 
k \cot (\delta) \, = \, -\frac{1}{a} \, + \, \frac{1}{2} r_e \,k^2 \, + \, v_2 \,  k^4 +
\, v_3 \, k^6 + \,v_4 \, k^8 + \ldots \label{effrangeexp}
\end{equation}
and is particularly useful in the case of unnaturally large $a$.
The $v_i$ coefficients at this order are fixed entirely by $m_\pi$ and $1/a$.
The low energy theorems were compared with $v_i$ extracted from a partial wave 
analysis of the scattering data.  All of the predictions were many times larger
than the $v_i$ extracted from scattering data.  As the low energy theorems are
particularly sensitive to pion physics (all terms are nonanalytic in $m_\pi$)
a plausible conclusion from this discrepancy is that the part of $Q$ counting
associated with the expansion of $m_\pi/\Lambda$
  has broken down.  Such a 
conclusion is consistent with the many successes of $Q$ counting\cite{KSW2} for deuteron properties provided such successes depend essentially on the expansion
of $1/(a \Lambda)$ rather than $m_\pi/\Lambda$.  Indeed, in ref. \cite{KSW2}
the authors show that the the effective range expansion without explicit pions
does a better job of describing the form factors at low momentum transfers
than the theory based on $Q$ counting with explicit pions.   This is precisely
what one would expect if the $1/(a \Lambda)$ expansion were working and the
$m_\pi/\Lambda$ failing. 
  
The scenario where $\Lambda$  is numerically of the same scale
 as $m_\pi$ is quite plausible.
The essence of $Q$ counting is that the {\it only} long distance scales are
$a$ and $1/m_\pi$.  As a practical matter one should identify $1/\Lambda$ as
the longest of the various short distance scales in the problem as that will be the scale responsible for a breakdown of the expansion.  The effective range,
 $r_e$, is
an important scale characterizing low energy nucleon-nucleon scattering.
Numerically it is $\sim 2.7$ fm in the singlet channel and $\sim 1.7$ fm in the
triplet channel.  In both cases, $r_e \, m_\pi > 1$.  If one identifies
$1/r_e$ as a short distance scale, $\Lambda$, then $m_\pi/\Lambda > 1$
and a chiral expansion is not valid.  Two issues must be resolved before coming to such a conclusion.  The first is whether $r_e$ is a ``short distance'' scale
(which just happens to be numerically long), and the second is whether the appropriate scale is $1/r_e$, or $1/r_e$ times some numerical factor which if large 
enough might render the chiral expansion useful.

The first issue can be easily resolved in the context of $Q$ counting.
If in $Q$ counting, $r_e$  were of order $Q^{-1}$, for example, scaling as $a$ or $1/m_\pi$, then the large numerical value of $r_e$ 
would be natural.  However,
the effective range has been calculated at leading nontrivial order
in $Q$ counting ({\it i.e.,} next to
 leading order) \cite{MS2} and it is explicitly seen that $r_e \sim Q^0$. Thus, the value of $r_e$ in the context of $Q$ counting
is determined by short distance scales.  This in turn suggests that the
longest scale treated as short distance in $Q$ counting (namely $r_e$)
is comparable to or larger than the shortest longest scale ($1/m_\pi$). The fact that $r_e m_\pi \ge 1$ suggests that the chiral expansion may not be under
 control even when the unnaturally large scattering length is taken into
 account. 

The issue of whether the large value of the effective range invalidates the chiral aspects of $Q$ counting is central to the effective field theory program
 in nuclear physics. 
The question of whether the low energy theorems of ref.~(\cite{CH2})
are badly violated
is, in turn, a critical issue in assessing the viability of
 the chiral aspects of the 
$Q$ counting scheme.  Recently, Mehen and Stewart\cite{MS1} have raised the 
question of whether  errors in the phase shifts render a reliable extraction of
the $v_i$ coefficients impossible. 
 They make a crude  estimate of the errors of the  $v_2$  in the triplet
 channel coefficient including the uncertainties 
using the  reported values from the Nijmegen partial wave analysis for 
$k <$ 70 MeV along with the scattering length and effective range and obtain
$v_2 =  -.50 \pm .52\, \pm \sim .1\, {\rm  fm}^3$, where the first error is 
a quadrature sum of the estimates errors and the second uncertainty is a theoretical  estimate of the contributions from the $v_3$ and higher terms.  This estimate is consistent 
 with  both the value fit from the Nijmegen analysis\cite{Nij}, $v_2^{\rm fit}
= .04 \, \, {\rm  fm}^3$, and the low energy theorem
 prediction value  of $ v_2^{\rm LET} =
.95 \,  \, {\rm  fm}^3$. 
 Using the second lowest report point from the Nijmegen analysis they estimate the $v_2 =  .03 \pm .04 \pm \sim .5 \, {\rm  fm}^3$.  Accordingly they conclude that there is too much uncertainty in the extraction of the $v_i$ coefficients
 to make a sharp test of the low energy theorems.

In this paper we will show that data are sufficiently good so that sharp
tests of the low energy theorems of ref.~\cite{CH2} are possible and that
the theorems are, in fact, badly violated.  The most sensitive method is
to consider weighted sums of the low energy theorems of the $v_i$
coefficients which can be extracted with far greater than precision than
the individual terms.  In particular, we test the total contribution to $k
\cot \delta$ arising from all of the higher terms ($v_2$ and above) in
the effective range expansion.  In this paper we will focus on tests in
the triplet channel.
 One particularly nice place to test is at the deuteron pole ($k = i
\sqrt{M B}$ where $B$ is the binding energy) which is known with great
precision.  One can also work at small real $k$ and compare with the
uncertainties in the partial wave analysis.  In this analysis, we find in
every case that the $Q$ counting at second order makes predictions which
are incompatible with the data.

\section{Tests of the low energy theorems}
For the following analysis it is useful to write an expression for
 $k \cot (\delta)$ excluding the contributions from the scattering length
and effective range terms.  We will refer to this quantity as the shape
function and denote it, ${ \cal S}(k^2)$.  Thus,
 \begin{eqnarray} {\cal
S}(k^2) \, &\equiv & \, k \cot (\delta) \, - \, \left( -\frac{1}{a} +
 \frac{1}{2} r_e k^2 \right )\\
&=& \sum_{j \ge 2} v_j k^{2 j}
\end{eqnarray}
where the second equality holds only within the radius of convergence of the
effective range expansion.  Note, however, that the general definition holds
for all $k^2$.  The shape function, ${\cal S}(k^2)$, can be calculated in the $Q$ expansion.  Using
the expression for $k \cot (\delta) $ in ref. \cite{CH2} which can be obtained
using either a cutoff scheme or dimensional regularization with either PDS or
OS subtraction, one finds at order $Q^2$ for either the singlet or triplet channel
 \begin{eqnarray}
{\cal S}^{\rm LET} (k^2)\,  & = & \,  \frac{g_A^2 M}{16 \pi^2 f_\pi} \,
\left [ \frac{1}{a^2} \, - \, \frac{2 m_\pi}{a} \, + 
k^2 \left ( - 1 \, + \frac{8}{3 m_\pi a} \, - \, 
\frac{2}{m_\pi^2 a^2} \right) \right] \nonumber \\ \nonumber\\
& - &\,  \frac{1}{a^2} \, \frac{g_A^2  M}{64 \pi f_\pi^2} \,
\left( \frac{m_\pi^2}{k^2} \right )
\ln \left
(1 + \frac{4 k^2}{m_\pi^2} \right ) \nonumber \\ \nonumber \\
 & + &  \, \frac{m_\pi}{a} \, \frac{g_A^2  M}{16 \pi f_\pi^2} \, 
 \left( \frac{m_\pi}{k} \right )\,
 \tan^{-1} \left ( \frac{2 k}{m_\pi} \right ) \,
 + \, m_\pi^2 \, \frac{g_A^2  M}{64 \pi f_\pi^2} \, 
 \ln \left (1 + \frac{4 k^2}{m_\pi^2} \right ) \, + {\cal O}(Q^3)
 \label{s}\end{eqnarray}

The predicted $v_i$ coefficients predicted  at  this
 order in the $Q$ expansion are     
obtained by differentiating the preceding expression with respect to $k$,
\begin{equation}
v_j \, = \, \frac{1}{(j-2)!} \,\left . \frac{\partial^2 {\cal S}^{\rm LET}}{\partial k^2} \right |_{k=0}
\end{equation}
 This gives
\begin{eqnarray}
v_2 \, &  = & \, \frac{g_A^2 M}{16 \pi f_\pi^2} \, \left ( \, -\frac{16}{3 a^2
\,
m_\pi^4}\,  + \, \frac{32}{5 a \,m_\pi^3} \, - \,\frac{2}{m_\pi^2} \right
)\nonumber \\ \nonumber \\
v_3 \, & = & \, \frac{g_A^2 M}{16 \pi  f_\pi^2} \, \left ( \, \frac{16}{ a^2 \,
m_\pi^6}\,  - \, \frac{128}{7 a \, m_\pi^5} \, + \,\frac{16}{3 m_\pi^4} \right )
\nonumber \\ \nonumber \\
v_4 \, & = & \,  \frac{g_A^2 M}{16 \pi  f_\pi^2} \, \left
( \, -\frac{256}{5 a^2 \, 
m_\pi^8}\,  + \, \frac{512}{9 a \,m_\pi^7} \, - \, \frac{16}{ m_\pi^6} \right )
 \nonumber \\ \nonumber \\ & \ldots & \label{vi} \end{eqnarray}
 The
effective range expansion from eq. (\ref{s}) has a finite radius of
convergence.  The existence of a cut at  $k^2 = - m_\pi^2/4$ implies that the
series only converges for $k^2 <m_\pi^2/4 \approx 70 {\rm MeV}$.  However
this limitation on region of the validity of the effective range expansion is
{\it not} a limitation on the range of validity of eq.~(\ref{s}).  It is
valid up to corrections of order $Q/\Lambda$ even for $k^2 > m_\pi/2$.  
Indeed, the entire motivation underlying the development of the $Q$
expansion was a scheme valid when $k \sim m_\pi$\cite{KSW}.

 Note that the prediction of the shape function ${\cal S}(k^2)$ depends on no
free parameters and thus is a low energy theorem in the same sense that
 the predictions for the $v_i$'s are low energy theorems. The low energy theorem for ${\cal S}(k)$ is more basic  than the low energy theorems for
 the $v_i$; all the predicted 
$v_i$ follow from eq.~(\ref{s}).   
It is also important to note that  ${\cal S}(k^2)$ at fixed $k^2$
is far easier
to extract from the data  with reliable error estimates than $v_i$ since all that is needed to be known is the phase shift, the scattering length and effective
range, along with knowledge of their errors.   One does not need to know enough information to accurately deduce higher derivatives of the function.
It should be noted, that within the radius of convergence of the effective range expansion,  {\it i.e.} for $k^2 < m_\pi^2/4$, testing the predicted
 ${\cal S}(k^2)$ tests a sum of the low energy theorems for the $v_j$ weighted by $k^{2j}$.
However, as noted above there is no necessity to restrict tests of
 ${\cal S}(k^2)$ to this regime.

It is important to note that the shape function, ${\cal S}(k^2)$, like the
individual $v_j$'s provide an ideal way to probe the pionic aspects of the $Q$
counting scheme.  Recall that in the $Q$ counting scheme there are
 two small mass scales
apart from the external momentum, $1/a$ and $m_\pi$.  However,  $1/a << m_\pi$.  Thus, it remains possible that  the underlying ``short distance'' scale,
$\Lambda$, is in fact  comparable to $m_\pi$  while $1/a  <<\Lambda$.  If such
a situation occurs one expects observables primarily sensitive to $1/(a \Lambda)$ to be 
well described, whereas observables primarily sensitive to $m_{\pi}/\Lambda$ to 
be poorly described.  We note that this possibility is not implausible given
experience with potential models which are fit to the data where it is generally seen that the non-one-pion-exchange part of the potential remains significant 
at ranges comparable to $1/m_\pi$ so that there is a ``short distance'' scale in the problem  of the pionic range\cite{CH} .  The most straightforward way to test whether the $m_\pi/\Lambda$ expansion is under control is to compare predictions
from a theory  with  pions integrated out to those which include pions and see
whether one gets systematic improvement by including the pions.  Unfortunately
for generic observables this test is not very clean since the observable may be
completely dominated  by the $1/(a \Lambda)$ expansion.  On the other hand, if
 one has an observable which vanishes at some order in the pion-integrated out
 theory but not in the pion-included theory than one has a prediction which
 explicitly tests the pionic contributions.  The shape function,
 ${\cal S}(k^2)$, at
order $Q^2$ is such an example (as are the $v_i$ coefficients derived from it).  The reason for this is that $k \cot {\delta}$ in the pion-integrated-out theory is  just the effective range expansion, which at order $Q^2$ truncates at the
second term and implies that ${\cal S}(k^2) =0$ at this order.  Thus the predictions
of ${\cal S}(k^2)$ provides a sharp test of the pionic part of $Q$ counting.

In this paper we will restrict our attention to the triplet channel as in that channel
 $a$ and $r_e$ have been extracted from the partial wave analysis
 with very small error bars\cite{Stoksa} allowing for a very sharp test.  They are given by
\begin{equation}  
a \, = \, 5.420 \pm .001 \, {\rm fm} \; \; \;
 r_e \, = \, 1.753  \pm .002 {\rm fm}
\end{equation}

An additional advantage to working in the triplet channel is the existence of 
the deuteron bound state which corresponds to a pole in the scattering amplitude when it is analytically continued to imaginary momentum.   Define the
 quantity,  $\gamma$, as
\begin{equation}
\gamma \, = \, \sqrt{M B}
\end{equation}
where $B$, the deuteron binding energy, is known with great precision to be 
B = 2.224575(9).   The  pole  occurs at $k = i \gamma $ and is
 fixed by the condition that denominator of the scattering amplitude vanishes.  This in turn  fixes the value of our shape function at 
$k^2=-\gamma^2 $:
\begin{equation}
{\cal S}(-\gamma^2) \, = \, \frac{1}{a} \, + \, \frac{1}{2} r_e \gamma^2 \,-  \, \gamma
\, = \, -.017  \pm .012 {\rm MeV}
\end{equation}
In contrast, the low energy theorem  gives --.743 MeV which deviates from the 
extracted value by more than 6 $\sigma$.  It is hard to argue that the discrepancy can be attributed to uncertainties in the data.

One can also test the low energy theorem for ${\cal S}(k^2)$ for real $k$.
We have used the values in the Nijmegen phase shift analysis \cite{Nij}.
The extraction of ${\cal S}(k^2)$ from the data involves subtracting the first two terms of the effective range expansion from the extracted $k \cot (\delta)$.
Since both of these quantities are intrinsically much larger 
 than ${\cal S}(k^2)$, it is essential for both  quantities 
to be given with as
much precision as possible.  In particular, one must be careful to use the
 relation between lab energy and center of mass momentum from ref. \cite{Nij}
which includes relativistic effects and the proton-neutron mass difference.  
In table~{\ref{tab1} we compare the extracted value of the shape function, 
${\cal S}(k^2)$, with the predictions from 
the low energy theorems. We have decided to include in this comparison
 lab energies up to  50 MeV,
corresponding to a momentum of 153 MeV, which is approximately  $m_\pi$.
As noted earlier, this highest energy is outside the radius of convergence of
 the effective range expansion ($k = m_\pi/2)$.
This is not a concern  if $\Lambda >> m_\pi$ (the assumption underlying the
chiral part of the  $Q$ expansion), since under this assumption
 $k \sim m_\pi$ is within the presumed domain of validity of
 the $Q^2$  expansion.  (We note that the points at the
deuteron  pole, and at positive energies of 1 and 5 MeV are within the
 radius of convergence of the effective range expansion.)
 In all cases except for ${\rm T_{lab} = 1~ MeV}$, we find that the low energy
theorem predicts  ${\cal S}$   significantly greater in magnitude
and the opposite sign from the extracted value.  (For
${\rm T_{lab} = 1~MeV}$, the low energy theorem presumably disagrees with the
sign of the actual value of ${\cal S}$ but the sign of the extracted
 value is undetermined since
the value is consistent with zero.)  The uncertainties associated 
 with the extraction are also given in the table.  The significant point
is that for all cases the disagreement between the extracted value and
the low energy theorem prediction is many standard deviations, even for the smallest values of $k$.  The
statistical significance of the discrepancy grows with $k$ so that by the top
of our energy range the predicted value differs from the extracted one
by more than 100 $\sigma$.
Clearly, the data has sufficient precision to test the low energy theorem and
it is equally clear that  the low energy theorem fails  to correctly predict 
${\cal S}$  even up to the  sign.

The   argument given above demonstrates conclusively that the $Q$
 counting scheme at second order  fails to predict the shape function
 ${\cal S}(q^2)$.
The central  purpose  of this paper is to  demonstrate explicitly that
 there exist
observables which are dominated by pionic physics for which $Q$ counting fails.
 For this purpose the  results discussed above is quite sufficient.  It is 
nevertheless of some interest to ask whether the data is good enough
to test the low energy theorems for the individual $v_j$ coefficients in 
the effective range expansion or whether, as suggested in ref. \cite{MS1}, the
 uncertainties are too large.  The difficulty of accurately extracting
high  derivatives of functions from data with uncertainties is clear. 

 Fortunately, there are effectively several distinct fits to the scattering
 data.  Note that the Nijmegen group not only fit the data directly
in their partial wave analysis \cite{Nij}, they also fit several potential
models directly to the data---{\it i.e., not the the partial wave analysis phase shifts} with a $\chi^2$ per degree of freedom of 1.03, essentially unity\cite{Nij2}. In effect, as noted in ref. \cite{Nij2}, the phase shifts produced by these
 potential models represent parameterizations of the partial wave analysis phase shifts.   Now, as discussed in ref. \cite{Stoksa} the effective range expansion coefficients $v_j$ extracted from the various potential models agree with 
each other and with the $v_i$ extracted directly from the best fit values of 
the partial wave analysis to an extremely high precision.  This is quite useful, since the bias introduced is presumably quite different in the the various 
fits.   Thus, overall the spread
 between the various potential models  and the direct fit to the partial wave
 analysis should provide some sense of the scale of the uncertainty.

In table ~\ref{tab2} we reproduce the triplet channel $v_j$ extracted from
 the partial wave analysis fits and from  the potential models and the values 
from the low energy theorems.  Note that for all cases the spread between the
different extracted values is quite small.  The largest relative spread is 
in the $v_2$ coefficient values and that is presumably because $v_2$ is
accidentally very   small.  In all cases, the spread in the values of the coefficients
is vastly smaller than the difference from any of these values to the one
 predicted  by the low energy theorems.  This strongly suggests that the 
individual $v_j$ coefficients are known well enough to test the low energy
 theorems and that low energy theorems make predictions inconsistent with the
data.

\section{Discussion}

By focusing on the quantity ${\cal S}(k^2)$, we have been able to show that
at least one pion aspect of $Q$  counting fails badly at next to leading order.
>From our analysis it is clear that if the uncertainty estimates of
 ref. \cite{Nij} are even approximately correct, then
 the predictions for 
 ${\cal S}(k^2)$ from the low energy theorems are in marked disagreement with the data, even to the point of getting the sign wrong.
One obvious explanation for this is the one advanced in our previous paper
\cite{CH2} and discussed in the introduction, namely that $1/m_\pi$ is not
long ranged compared to other scales in the problem.  This possibility
is plausible on its face, since it is  known in nuclear physics that there
are many length scales which are comparable to $1/m_\pi$ but which have no
obvious chiral origin.  The effective range is a good example.  Another
example is the characteristic ranges of the non-pion-exchange part 
of nuclear potential which are fit to phase shifts (although as discussed in
 ref.  \cite{Scal} the need to fit the effective range constrained the non-pionic
part of the potential to be long).  While  this does not prove that the pionic
part of $Q$ counting must fail, it certainly makes it very plausible.

If the failure of the low energy theorems for scattering indicates a systematic
failure of pionic effects of s-wave properties in $Q$ counting, one expects
 failure for other observables in the sense that the explicit inclusion of pions should not lead to improved predictive power.
 The recent calculations
of deuteron form factors in ref. \cite{KSW2} strongly support this view.
The calculation of the form factors using a simple  effective range expansion
treatment including up to the effective range describes the data better
than the next-to-leading order treatment including explicit pions.  Had
the pionic aspects of $Q$ counting been under control one would have expected
the calculation including explicit pions would have improved things.

At present we know of no observable associated with s-wave two nucleon states
for which the inclusion of explicit pions in $Q$ counting improves predictions
and several for which it worsens them.  Of course, this does not prove that
the $m_\pi/\Lambda$ expansion will generally fail for all s-wave observables.  
It remains possible, for
example,
 that one coefficient in the next-to-leading order theory is accidentally
large and that by fitting it and working at next-to-next-to leading order
the usefulness of the $m_\pi/\Lambda$ will be manifest.  
The authors of ref. \cite{KSW2} assert (without proof) that at higher orders 
the effective field theory with pions will work better than the simple effective range calculation since it has the correct underlying physics.
We believe that this scenrio is unikely in view of the fact that there seems
to be no  scale separation between 1/$m_\pi$ and
 ``short distance'' scales. 

It is clear how to test this idea: calculate observables at higher order for 
theories with and without explicit pions and compare the qualities of the
prediction.  In doing such comparisons, however, it is essential to distinguish
the quality of the descriptions of the underlying physics from the quality of
mere curve fitting.  Accordingly, in such comparisons it is essential that the
theories with the same number of parameters be compared and the same prescriptions for fitting.   Thus, for example the appropriate test for the deuteron form factors of ref. \cite{KSW2} is not whether a higher order effective field theory calculation wit
h pions out performs the simple effective range expansion---eventually it must, at least over some region, as one will have additional parameters to characterize the current operator.  Rather, the test is whether the theroy with pions out performs an effe
ctive field theory  with pions
integrated out and with  the same number of parameters.

The authors thank Silas Beane and Daniel Phillips for interesting
discussions. 
TDC gratefully acknowledges the support
of the U.S. Department of Energy under grant no. DE-FG02-93ER-40762.

\newpage

\begin{table}[p]
\begin{tabular}{||c  |c | c||}
lab energy (MeV) &${\cal S}$ extracted (Mev)
& ${\cal S}$ low energy theorem (Mev)\\
\hline \hline
Deuteron Pole &  $-0.0017 \pm  0.0125$ & -0.743\\
1  &  $-0.00095 \pm 0.00721$ & -0.0258 \\
5  & $0.0428 \pm 0.0194$ & -0.535 \\
10 & $0.245 \pm 0.047$  &  -1.78 \\
25 &  $ 2.18 \pm 0.14$ & -7.54 \\
50 & $  11.03 \pm 0.24 $ & -20.10\\
\hline \hline
\end{tabular}
\caption{A comparison of the shape function, ${\cal S}(k^2) =
k \cot (\delta) + 1/a - 1/2 r_e k^2 $ for the ${}^3 S_1$ channel extracted
from the Nijmegen partial wave analysis with the prediction by the low energy theorem of eq. ~(\ref{s})}
\label{tab1}
\end{table}\vspace{.5 in}

\begin{table}[p]

\begin{tabular}{||c| c | c | c ||}
$\delta$ (${}^3S_1$  channel)&$v_2$ (${\rm fm}^{3})$ & $v_3$ (${\rm fm}^{5})$ & $v_4$ (${\rm
fm}^{7})$\\ \hline \hline & & & \\
low energy theorem &  -.95 & 4.6  & -25.  \\  \hline & & & \\
partial wave analysis & .040 & .672 &  -3.96\\ 
Nijm I                & .046 & .675 & -3.97 \\
Nijm  II              & .045 & .673 & -3.95 \\
Reid93                & 0.33 & .671 & -3.90 \\
 \end{tabular}

\caption{A comparison of the effective range expansion 
coefficients, $v_i$,   for the ${}^3S_1$ 
and ${}^3S_1$ 
channel predicted from the low energy theorem with coefficients extracted
from the  partial wave analysis and with three potential models---Nijmegen I, Nijmegen II and Reid 93---which were 
 fit directly to the scattering data.}
\label{tab2}
\end{table}

\end{document}